\documentclass[twocolumn,showpacs,showkeys,amsmath,amssymb]{revtex4}
\usepackage{graphicx}
\def\eV{\hbox{ eV}}
\def\GeV{\hbox{ GeV}}
\begin{document}
\title{Neutrino mass in GUT constrained supersymmetry with 
       R-parity violation \\ in light of neutrino oscillations}
\author{Marek G\'o\'zd\'z} 
\email{mgozdz@kft.umcs.lublin.pl}
\author{Wies{\l}aw A. Kami\'nski}
\email{kaminski@neuron.umcs.lublin.pl} 
\affiliation{Department of Theoretical Physics, Maria
Curie--Sk{\l}odowska University, Lublin, Poland}
\author{Fedor \v Simkovic} 
\email{simkovic@fmph.uniba.sk}
\affiliation{Department of Nuclear Physics, Comenius University,
Bratislava, Slovakia}
\begin{abstract}
The neutrino masses are generated in grand unified theory (GUT)
constrained supersymmetric model with R-parity violation. The neutrinos
acquire masses via tree-level neutrino-neutralino mixing as well as via
one-loop radiative corrections. The theoretical mass matrix is compared
with the phenomenological one, which is reconstructed by using neutrino
oscillation and neutrinoless double beta decay data. This procedure
allows to obtain significantly stronger constraints on R-parity breaking
parameters than those existing in the literature. The implication of
normal and inverted neutrino mass hierarchy on the sneutrino expectation
values, lepton-Higgs bilinear and trilinear R-parity breaking couplings
is also discussed.
\end{abstract}


\maketitle 

\section{Introduction} 

The discovery of neutrino oscillations triggered a natural interest in
the problem of masses of neutrinos. Unfortunately, in the oscillation
experiments only the differences of the squares of masses can be
determined. The importance of knowing the absolute scale of neutrino
masses is obvious. This knowledge will allow to set direction in which
the changes of the standard model of particles and interactions should
go; what is more, the problems of dark matter and dark energy, lepto-
and baryogenesis, evolution of the Universe and many other could be
addressed. Even if the values of neutrino masses will be measured in
experiments, the question about the mechanism of obtaining those masses
remains open, just in the same way as the widely approved Higgs
mechanism is still not experimentally confirmed.

There are many proposals of generating the neutrino mass
matrix. Starting from ad-hoc ans\"atze, through the most-widely approved
see-saw model, through extra dimensions, through a result of
supersymmetry breaking. Among these one also finds the loop mechanism
for Majorana neutrinos. The effective vertex of the form $\bar\nu\nu$ is
there expanded to contain a squark-quark or slepton-lepton loop.  This
setting introduces the R-parity violation, therefore it needs to be
described within a supersymmetric model with explicitly or spontaneously
broken R-parity. Such models provide an elegant way of not only
resolving the naturally small neutrino mass problem, but also introduce
supersymmetry, needed by the string theory, solve the hierarchy problem,
provide much better description of the anomalous magnetic moment of the
muon and much more.

The problem of R-parity violation in supersymmetric models received a
great deal of attention during last few years \cite{RpV}. The studies of
this topic were connected with, among others, leptonic decays
\cite{dreiner}, gravitino decays \cite{moreau} and the problem of
neutrino masses and oscillations \cite{bedny,nu-RpV,abada}. In general,
supersymmetric models with R-parity violation (RpV) fall into one of the
three categories. First, we have the spontaneous RpV. In this case the
R-parity is violated by a non-zero vacuum expectation value of some
scalar field \cite{spon-RpV}. Another possibility is the explicit
breaking of R-parity by introduction of bilinear and/or trilinear
terms. The bilinear RpV models \cite{2RpV} are characterized by good
predictivity due to a small number of parameters. The third category is
the explicit RpV by trilinear terms present in the superpotential
\cite{3RpV,Haug}. In this case one allows for the presence of bilinear
terms, since these would anyway show up during the RGE evolution of
trilinear coupling constants, assuming at the same time that they do not
affect the phenomenology of the trilinear terms. This is motivated by
the fact that there is no fine tuning among different contributions,
which can therefore be analyzed separately. The trilinear scenarios are
the most studied due to the reachest phenomenology and possibility of
obtaining most interesting limits on non-standard physics parameters.

Our work follows the line of research concerning the generation of
neutrino masses in SUSY without R-parity, that has been developed in the
last few years. The aim of our paper is to get new individual limits on
the R-parity breaking parameters by taking the advantage of the recent
data on neutrino oscillations and neutrinoless double beta decay. In
addition, the previous studies presented in \cite{abada,Haug,Bhatta} are
improved by a more accurate treatment of the neutrino mass
contributions, in particular by reducing the dependence on the SUSY
parameter space. The considered model is the minimal supersymmetric
standard model with supersymmetry breaking transmitted by (super)gravity
interactions (SUGRA MSSM) \cite{mssm}, with the squark and slepton
mixing phenomena properly included. The RGE evolution within GUT
constrained SUGRA MSSM is introduced to obtain the low-energy particle
spectrum. The GUT constraints involve unifying masses and coupling
constants to some common values at the GUT scale. These are universal
scalar mass $m_0$, gaugino mass $m_{1/2}$, trilinear scalar coupling
$A_0$, the ratio of the Higgs vacuum expectation values $\tan\beta$ and
the sign of the bilinear Higgs mixing parameter $\mu$.

The paper is organized as follows. In the next section we present the
model and describe our procedure of finding low energy spectrum of SUSY
particles using the GUT constraints. We present also improved versions
of contributions of different loops to the neutrino mass matrix. In
Sec. III we present upper limits on various combinations of RpV coupling
constants, both the trilinear $\lambda$'s and dimensionful bilinears
$\Lambda$'s. Discussion and conclusions follow at the end.

\section{The model}

The loop mass mechanism may be described in the framework of R-parity
violating MSSM (RpVMSSM) with trilinear and bilinear soft breaking
terms. The MSSM (see e.g. \cite{mssm}) and its many variations are well
known in the literature. Here, we closely follow Ref.\
\cite{HirschValle} regarding the conventions and construction of the
mass matrices. In short, RpVMSSM is characterized by the superpotential
which consists of the R-parity conserving part
\begin{eqnarray}
W^{MSSM} &=& \epsilon_{ab} [(\mathbf{Y}_E)_{ij} L_i^a H_1^b \bar E_j
+ (\mathbf{Y}_D)_{ij} Q_i^{ax} H_1^b \bar D_{jx} \nonumber \\
&+& (\mathbf{Y}_U)_{ij} Q_i^{ax} H_2^b \bar U_{jx} + \mu H_1^a H_2^b ],
\label{wmssm}
\end{eqnarray}
and the R-parity violating part
\begin{eqnarray}
W^{\not R_p} &=& \epsilon_{ab}\left[
\frac{1}{2} \lambda_{ijk} L_i^a L_j^b \bar E_k
+ \lambda'_{ijk} L_i^a Q_j^{xb} \bar D_{kx} \right] \nonumber \\
&+& \frac{1}{2}\epsilon_{xyz} \lambda''_{ijk}\bar U_i^x\bar
D_j^y \bar D_k^z + \epsilon_{ab}\kappa^i L_i^a H_2^b.
\end{eqnarray}
The {\bf Y}'s are 3$\times$3 Yukawa matrices. $L$ and $Q$ are the $SU(2)$
left-handed doublets while $\bar E$, $\bar U$ and $\bar D$ denote the
right-handed lepton, up-quark and down-quark $SU(2)$ singlets,
respectively. $H_1$ and $H_2$ mean two Higgs doublets. We have introduced
color indices $x,y,z = 1,2,3$, generation indices $i,j,k=1,2,3$ and the SU(2)
spinor indices $a,b,c = 1,2$. In order to get rid of too rapid proton decay
and to describe lepton number violating processes, like the neutrinoless
double beta decay, it is customary to set $\lambda''=0$.

We supply the model with scalar mass term
\begin{eqnarray}
{\cal L}^{mass} &=& \mathbf{m}^2_{H_1} h_1^\dagger h_1 +
                    \mathbf{m}^2_{H_2} h_2^\dagger h_2 +
     q^\dagger \mathbf {m}^2_Q q + l^\dagger \mathbf {m}^2_L l \nonumber \\
&+&  u \mathbf {m}^2_U u^\dagger + d \mathbf {m}^2_D d^\dagger +
     e \mathbf {m}^2_E e^\dagger,
\end{eqnarray}
soft gauginos mass term
\begin{equation}
  {\cal L}^{gaug.} = \frac12 \left( 
  M_1 \tilde{B}^\dagger \tilde{B} + 
  M_2 \tilde{W_i}^\dagger \tilde{W^i} +
  M_3 \tilde{g_a}^\dagger \tilde{g^a} + h.c.\right ),
\end{equation}
as well as the supergravity mechanism of supersymmetry breaking, by
introducing the Lagrangian
\begin{eqnarray}
{\cal L}^{soft} &=& \epsilon_{ab} [(\mathbf{A}_E)_{ij} l_i^a h_1^b \bar e_j
+ (\mathbf{A}_D)_{ij} q_i^{ax} h_1^b \bar d_{jx} \nonumber \\
&+& (\mathbf{A}_U)_{ij} q_i^{ax} h_2^b \bar u_{jx} + B \mu h_1^a h_2^b +
B_2 \epsilon_i l_i^a h_2^b],
\end{eqnarray}
where lowercase letters stand for scalar components of respective chiral
superfields, and 3$\times$3 matrices {\bf A} as well as $B\mu$ an $B_2$ are
the soft breaking coupling constants.

All the running parameters are obtained by using the renormalization
group equations (RGE) \cite{wodprc99,mg-gmsb}. At the beginning, one
evolves all gauge and Yukawa couplings for three generations up to the
GUT scale $M_{GUT}\sim 10^{16} \GeV$. We use the one-loop standard model
RGE \cite{drtjones} below the mass threshold, where SUSY particles start
to contribute, and the MSSM RGE \cite{martinvaughn} above that
scale. The contribution of 2-loop diagrams as well as those coming from
the R-parity violating couplings has been proven to be irrelevant in
discussions such as ours \cite{mg-art1}. The SUSY scale is initially set
to 1 TeV for all particles and is dynamically modified together with
evolution of their masses. At the GUT scale the masses of all scalars
and fermions are set to a common value $m_0 = m_{1/2}=m$. We have
considered a ``small'', chosen to be 150 GeV, and a ``large'' (1000 GeV)
value of $m$ in our analysis. We unify also the soft trilinear couplings
according to $\mathbf{A}_i = A_0 \mathbf{Y}_i$, with $A_0=500\GeV$. We
postpone the discussion of the influence of $A_0$ on the results to a
forthcoming paper. In the next step we construct all the relevant mass
matrices (squark, slepton, chargino and neutralino) and perform RGE
evolution of all the quantities back to $M_Z$ scale, taking care of the
minimization of the tree-level Higgs potential (important for EWSB
breaking) and radiative corrections. After iterating this procedure and
obtaining stable values of the parameters, we confront the obtained
values with restrictions coming from the present theoretical assumptions
and phenomenological data. Those constraints involve (1) finite values
of Yukawa couplings at the GUT scale; (2) proper treatment of
electroweak symmetry breaking; (3) requirement of physically acceptable
mass eigenvalues at low energies; (4) FCNC phenomenology.

The first problem is related to the values of $\tan\beta$ and is checked
during the RGE procedure. For very small $\tan\beta$ $(< 1.8)$ the top
Yukawa coupling may ``explode'' before reaching the GUT scale. It
follows from the fact that $Y_{top}(M_Z) \sim 1/\sin \beta$. Similarly,
other couplings $Y_{b}$ and $Y_\tau$ ``blow up'' before the GUT scale
for $\tan\beta > 50$ because they are proportional to $ 1/\cos \beta$ at
electroweak scale. In our analysis we have kept $\tan\beta \approx 20$
leaving the detailed discussion to a forthcoming paper.

Another theoretical constraint is imposed by the EWSB mechanism. In
order to obtain a stable minimum of the scalar potential, the following
conditions must hold:
\begin{eqnarray}
  (\mu B)^2 &>& ( \left|\mu\right|^2 + m_{H_u}^2 )
                ( \left|\mu\right|^2 + m_{H_d}^2 ), \nonumber \\
  2B\mu &<& 2 \left|\mu\right|^2 + m_{H_u}^2 + m_{H_d}^2.
\label{EWSB}
\end{eqnarray}
They are always checked in our procedure during RGE running, and points
which do not fulfill these conditions are rejected. Next restriction
comes from the requirement of positive eigenvalues of mass matrices
squared at the electroweak scale. The last requirement (see, e.g.,
\cite{mg-gmsb} for details) comes from the strongly experimentally
suppressed FCNC processes and provides the most severe constraints.

The so-obtained low energy spectrum is then used in further
calculations. Although, as will be seen, only the squarks and sleptons
masses enter the formulas, they depend in a complicated way on all other
masses and coupling constants through RGE equations
\cite{martinvaughn}. Therefore a complete and careful treatment is
necessary.

The neutrino mass matrix consists in our approach of three main parts:
\begin{equation}
  {\cal M}_\nu = {\cal M}^{tree} + {\cal M}^l + {\cal M}^q,
\label{Mnu}
\end{equation}
which are the tree level value and the contributions coming from
lepton-slepton and quark-squark loops, respectively. We note that there
are other terms that may be included in Eq. (\ref{Mnu}), in particular
loops contributions with bilinear insertions \cite{nu-RpV}. For the sake
of simplicity, however, we do not consider them by following the popular
approach to get individual limits on R-parity breaking parameters. We
adopt the conventional hypothesis that different contributions do not
significantly compensate each other and for this reason it is possible
to extract limits on individual contributions without knowing the
others.

Let us first recall the well known results. In the lowest order, the
contribution to the mass matrix reads \cite{Haug}
\begin{eqnarray}
  && {\cal M}^{tree}_{ii'} = \Lambda_i \Lambda_{i'} \ g_2^2 \cr
  &&\times  \frac{M_1 + M_2 \tan^2\theta_W}
  {4(\mu M_W^2 (M_1 + M_2 \tan^2\theta_W)\sin2\beta - M_1 M_2\mu^2)},
  \cr &
\label{Mtree}
\end{eqnarray}
where $\Lambda_i = \mu \langle \tilde{\nu}_i \rangle - \langle H_1
\rangle \kappa_i$, and $\langle \tilde{\nu}_i \rangle$ are the vacuum
expectation values of the sneutrino fields.

\begin{figure}
 \includegraphics[width=0.4\textwidth]{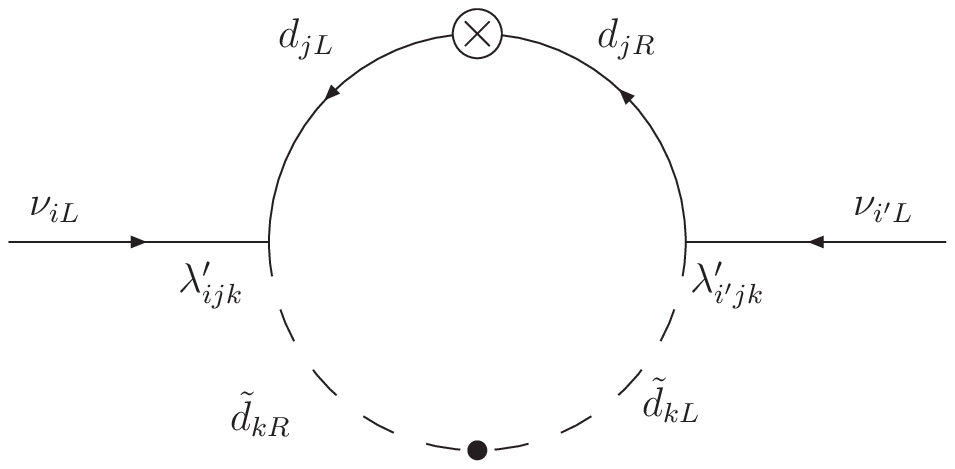}
 \includegraphics[width=0.4\textwidth]{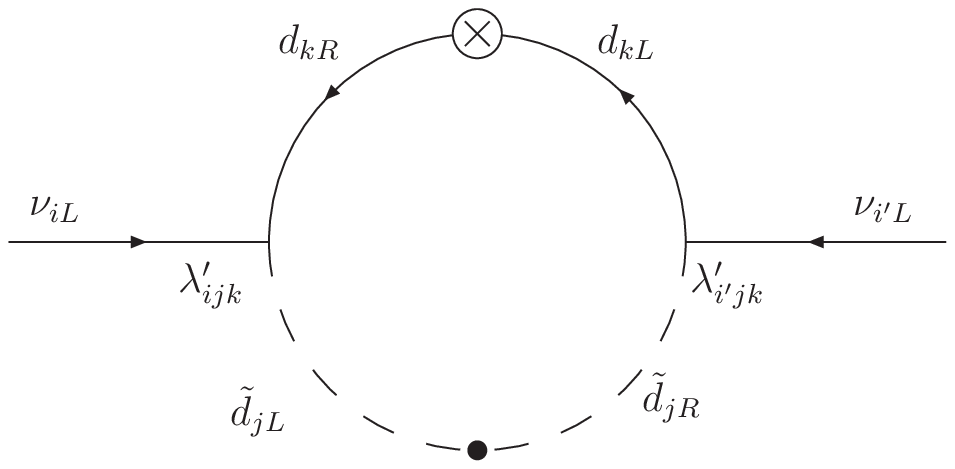}
 \caption{\label{Fig1} Feynman diagrams representing the squark-quark
 loop contribution to the Majorana neutrino mass.}
\end{figure}

Beyond the tree-level, the Majorana neutrino mass matrix may also be
generated by considering one loop self-energy diagrams. The particles
that propagate inside the loops are either quark and squark or lepton
and slepton. Let us start with the squark-quark loop. The relevant
Feynman diagrams are shown on Fig.\ \ref{Fig1}. It is important to note,
that one may (and should, if one wants to be very accurate) consider not
only the trilinear couplings, but also the mass insertions described by
the bilinear terms in the superpotential and Lagrangian (see
e.g. \cite{2RpV}). Here, however, we take the conventional approach and
assume that the phenomenology of the trilinear contribution (which is
more interesting from the point of view of obtaining constraints on the
RpV coupling constants) remains unaffected by the contribution coming
from bilinear terms.
%
\begin{figure}
 \includegraphics[width=0.4\textwidth]{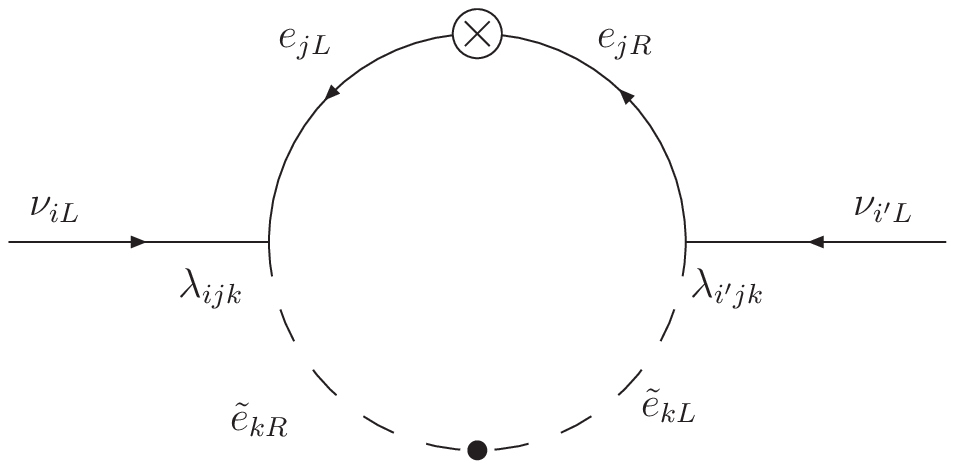}
 \includegraphics[width=0.4\textwidth]{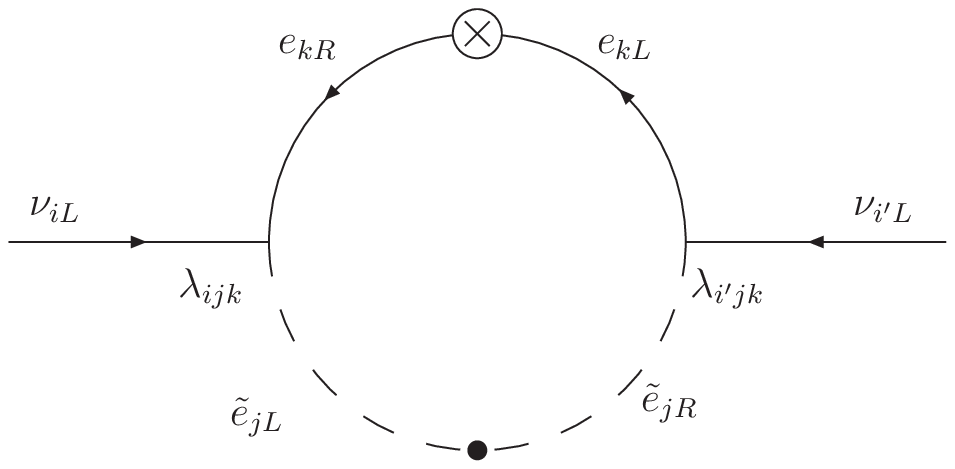}
 \caption{\label{Fig2} Feynman diagrams representing the slepton-lepton
 loop contribution to the Majorana neutrino mass.}
\end{figure}
%
Contrary to earlier approaches \cite{Haug,Bhatta}, we take into account
the down-squark mixing exactly
\begin{eqnarray}
 \tilde d_L &=& \phantom{-}\tilde d_1 \cos\theta + \tilde d_2 \sin\theta, \cr
 \tilde d_R &=& -\tilde d_1 \sin\theta + \tilde d_2 \cos\theta.
\end{eqnarray}
Here $L$ and $R$ label the left- and right-handed squark states in the
weak basis, while the $1$ and $2$ subscripts denote the two mass
eigenstates. The mixing angle is defined by
\begin{eqnarray}
  \sin(2\theta^k) &=& 2 m_{q^k} (A_k + \mu \tan\beta) \cr
  &\times& [(m_{\tilde q_L^k}^2 - m_{\tilde q_R^k}^2 - 0.34 M_Z^2
    \cos(2\beta))^2 \cr
  &-& 4 m_{q^k} (A_k + \mu \tan\beta)]^{-1/2}
\end{eqnarray}
with $A_k=(\mathbf{A}_D)_{kk}$ and $\tan\beta$ being the ratio of Higgs
vacuum expectation values. The squark mass eigenstates take the forms
\begin{eqnarray}
  m_{\tilde q_1^j}^2 &=&
    \frac12 (m_{\tilde q_L^j}^2 + m_{\tilde q_R^j}^2) + m_{q^j} \Big ( 
    m_{q^j} - \frac{A_j + \mu \tan\beta}{\sin(2\theta^j)} \Big) \cr
    &-& \frac14 M_Z \cos(2\beta), 
\cr
  m_{\tilde q_2^j}^2 &=& 
    \frac12 (m_{\tilde q_L^j}^2 + m_{\tilde q_R^j}^2) + m_{q^j} \Big ( 
    m_{q^j} + \frac{A_j + \mu \tan\beta}{\sin(2\theta^j)} \Big) \cr
    &-& \frac14 M_Z \cos(2\beta).
\end{eqnarray}
By introducing two dimensionless quantities, $x_1^{jk} \equiv m_{q^j}^2
/ m_{\tilde q_1^k}^2$ and $x_2^{jk} \equiv m_{q^j}^2 / m_{\tilde
q_2^k}^2$, one arrives at the following form of the neutrino mass matrix
\begin{eqnarray}
  {\cal M}_{ii'}^q &=& \frac{3}{16\pi^2}
  \lambda'_{ijk} \lambda'_{i'kj} \Big[ 
    \sin(2\theta^k) m_{q^j} \cr
  &\times& \left( \frac{\log(x_2^{jk})}{x_2^{jk} - 1} +
   \frac{(x_2^{jk} + 1) \log(x_1^{jk})}{(x_1^{jk} - 1)(x_2^{jk} - 1)}
   \right) \cr
  &+& (j\leftrightarrow k) \Big].
\label{Mqq}
\end{eqnarray}

We may repeat the same calculation for the slepton-lepton loop (see
Fig.\ \ref{Fig2}), just replacing in all definitions the squark masses
and mixing by analogous quantities for sleptons, as well as the quark
masses $m_{q^j}$ with lepton masses $m_{e^j}$. The only difference will
be the lack of the factor 3, which came in the previous case from
summing over the three colors of quarks and, of course, different
coupling constants. We end up with
\begin{eqnarray}
  {\cal M}_{ii'}^l &=& \frac{1}{16\pi^2}
  \lambda_{ijk} \lambda_{i'kj} \Big[ 
    \sin(2\phi^k) m_{e^j} \cr
  &\times& \left( \frac{\log(y_2^{jk})}{y_2^{jk} - 1} +
   \frac{(y_2^{jk} + 1) \log(y_1^{jk})}{(y_1^{jk} - 1)(y_2^{jk} - 1)}
   \right) \cr
  &+& (j\leftrightarrow k) \Big], 
\label{Mll}
\end{eqnarray}
where now $\phi$ is the slepton mixing angle, $y_1^{jk} \equiv m_{e^j}^2
/ m_{\tilde l_1^k}^2$ and $y_2^{jk} \equiv m_{e^j}^2 / m_{\tilde
l_2^k}^2$.

Let us now explain the procedure for finding constraints on the various
products of coupling constants $\lambda$, $\lambda'$, and $\Lambda$. The
right hand sides of Eqs. (\ref{Mtree}), (\ref{Mqq}), and (\ref{Mll}) can
be calculated from the MSSM RGE runnigs, during which the low energy
particle/sparticle spectrum is generated. We use random scatter to find
sets of physically relevant values of the various parameters. In the
next step the theoretical neutrino mass matrix is compared with the
phenomenological three neutrino mass matrix in the flavor space, which
is connected to the physical neutrino masses $m_i$ by the mixing matrix
$U$ through the relation $ {\cal M}^{ph} = U\cdot diag(m_1,m_2,m_3)\cdot
U^T$. The standard parameterization of the
Pontecorvo-Maki-Nakagawa-Sakata (PMNS) unitary matrix $U$ in terms of
the three angles is
\begin{eqnarray}
  \left (
  \begin{array}{ccc}
    c_{12} c_{13} & s_{12} c_{13} & s_{13} e^{-i \delta} \\
    -s_{12} c_{23} - c_{12} s_{23} s_{13} & c_{12} c_{23} - s_{12}
    s_{23} s_{13} & s_{23} c_{13} \\
    s_{12} s_{23} - c_{12} c_{23} s_{13} & -c_{12} s_{23} - s_{12}
    c_{23} s_{13} & c_{23} c_{13}
  \end{array}
  \right ) \nonumber\\
\times  \left (
  \begin{array}{ccc}
    1 & 0 & 0 \\
    0 & e^{i \alpha_{21}} & 0 \\
    0 & 0 & e^{i \alpha_{31}} 
  \end{array}
  \right ),~~~~~~~~~~~~~~~
\end{eqnarray}
where $s_{ij} = \sin\theta_{ij}$, $c_{ij} = \cos\theta_{ij}$, and
$\theta_{ij}$ is the mixing angle between the flavor eigenstates labeled
by indices $i$ and $j$. The recent global analysis of neutrino
oscillations \cite{maltoni} yields the best fit values: $
{\sin}^2\theta_{12}~=~0.3$, ${\sin}^2\theta_{23}~=~0.5$ and $
{\sin}^2\theta_{13}~=~0.002$. Note that for Majorana particles there
appear three CP violating phases, one Dirac phase $\delta$ and two
Majorana phases ($\alpha_{21}$ and $\alpha_{31}$), which remain
undetermined. Assuming the CP phases to be negligible one gets
\begin{equation}
  U= \left (
  \begin{array}{ccc}
    0.83 &  0.55 & 0.07 \\
   -0.42 &  0.55 & 0.72 \\
    0.35 & -0.63 & 0.69 
  \end{array}
  \right ).
\label{U}
\end{equation}

The absolute scale of neutrino masses is not determined by the neutrino
oscillations, which depend only on differences of masses squared. From
the global analysis \cite{maltoni} of neutrino oscillations the best fit
values $\Delta m^2_{21}~=~6.9~10^{-5}~\eV^2$ and $\Delta
m^2_{31}~=~2.3~10^{-3}~\eV^2$ are known. The three possible neutrino
mass patterns are frequently considered \cite{bfs}:\\ i) {\it The normal
hierarchy (NH) of neutrino masses}, which correspond to the case $m_{1}
\ll m_{2} \ll m_{3}$. Then we have $m_{1}\ll \sqrt{ \Delta
m^{2}_{{21}}}$, $m_{2}\simeq \sqrt{ \Delta m^{2}_{{21}}}$ and
$m_{3}\simeq \sqrt{ \Delta m^{2}_{{31}}}$.\\ ii) {\em Inverted hierarchy
(IH) of neutrino masses.} It is given by the condition $m_{3} \ll m_{1}
< m_{2}$. In the case for neutrino masses we have $m_{3} \ll \sqrt{
\Delta m^{2}_{\rm{31}}}$ and $m_{1}\simeq m_2 \simeq \sqrt{ \Delta
m^{2}_{\rm{31}}}$.\\ iii) {\it Almost degenerate neutrino mass
spectrum:} $m_{1} \simeq m_{2} \simeq m_{3}$. This case does not exclude
the possibility that the lightest neutrino is much larger than $\sqrt{
\Delta m^{2}_{\rm{31}}}$.

The absolute scale of neutrino masses can be determined by the
observation of the end-point part of the electron spectrum of Tritium
$\beta$-decay, the observation of large-scale structures in the early
universe and the detection of the neutrinoless double beta decay
($0\nu\beta\beta$-decay), if neutrinos are Majorana particles. The
amplitude of the $0\nu\beta\beta$-decay is proportional to the effective
Majorana neutrino mass $m_{\beta \beta} = U^{2}_{e1}\,m_{1} +
U^{2}_{e2}\,m_{2} +U^{2}_{e3}\,m_{3}$. This process has not been seen
experimentally until now and the best results have been achieved in the
Heidelberg-Moscow (H-M) experiment ($T^{0\nu}_{1/2} \ge 1.9\times
10^{25}$~y) \cite{H-M}. (Recently, some authors of the H-M collaboration
have claimed the experimental observation of the $0\nu\beta\beta$-decay
of $^{76}$Ge \cite{evid}. But the Moscow participants of the H-M
collaboration, performing a separate analysis of the data, found no
indication in favor of the evidence of the $0\nu\beta\beta$-decay
\cite{nanp}. The disproof or the confirmation of the claim will come
from future experiments.) By assuming the nuclear matrix element of
Ref. \cite{FedorVogel} we end up with $|m_{\beta\beta}| \le
0.55~\eV$. With this additional input limit we can find the maximal
allowed values for the matrix elements ${\cal M}^{ph}_{ij}$ of the
neutrino mass matrix, which are as follows:
\begin{eqnarray}
|{\cal M}^{ph-HM}|= 
  \left (
  \begin{array}{ccc}
    0.55 & 0.71 & 0.70 \\
    0.71 & 0.65 & 0.70 \\
    0.70 & 0.70 & 0.76 
  \end{array}
  \right )~ \eV.
\label{M_HM}
\end{eqnarray}
The elements of this matrix were obtained by assuming the whole allowed
mass parameter space of neutrinos and all possible CP-phases of the
neutrino mass eigenstates \cite{Haug}. In the calculation we used the
best-fit values of neutrino oscillation parameters given in
Ref. \cite{maltoni}. The elements of matrix (\ref{M_HM}) can be used to
test various theoretical approaches and allows one to extract limits on
certain fundamental parameters. Of course, one can not expect that by
diagonalizing of this matrix a relevant information on the masses of
neutrinos is obtained as each element of this matrix is a result of
analysis of all possible mixing of three neutrinos allowed by the
neutrino oscillations and the $0\nu\beta\beta$-decay data.

Instead of taking into account the current limit on $m_{\beta\beta}$
identified with the element ${\cal M}^{ph}_{ee}$ we consider also other
scenarios by assuming that the normal or inverted hierarchy of neutrino
masses is realized in the nature. Then we get
\begin{eqnarray*}
  && |{\cal M}^{ph-NH}| = 10^{-4}~\eV\cr 
  &&\times \left (
  \begin{array}{ccc}
    (22.4-27.2) & (0.64-49.4) & (5.16-52.0) \\
    (0.64-49.4) & (223-273)   & (210-267)   \\
    (5.16-52.0) & (210-267)   & (196-262)
  \end{array}
  \right ),
\end{eqnarray*}
\begin{eqnarray*}
  && |{\cal M}^{ph-IH}| = 10^{-2}~\eV\cr
  &&\times \left (
  \begin{array}{ccc}
    (1.86-4.72) & (0.22-3.11) & (0.26-3.05) \\
    (0.22-3.11) & (0.62-2.30) & (0.96-2.37) \\
    (0.26-3.05) & (0.96-2.37) & (1.31-2.50) 
  \end{array}
  \right ).
\end{eqnarray*}
These neutrino mass matrices were calculated by the assumption that the
mass of the lightest neutrino is negligible (see the above definitions
of the NH and the IH of neutrino masses).

\section{Results and Conclusions}

By confronting the phenomenological neutrino mass matrix ${\cal
M}^{ph-HM}$, derived from the analysis of the neutrino data, with the
theoretical mass matrix calculated within the R-parity breaking MSSM, it
is possible to find constraints on various combinations of the lepton
number violating $\lambda$, $\lambda'$ and $\Lambda$ couplings, which
enter Eqs. (\ref{Mtree}), (\ref{Mqq}), and (\ref{Mll}). If ${\cal
M}^{ph-NH}$ and ${\cal M}^{ph-IH}$ neutrino mass matrices are confronted
with the theory, one ends up with predictions for the R-parity violation
couplings. By considering the maximal values of these matrices the
largest possible values of R-parity breaking parameters are obtained.
We note that the predictions for R-parity breaking mechanisms associated
with normal or inverted mass hierarchy are deduced by the assumption
that one given mechanism dominates at a time. However, this scenario
might be excluded by other phenomenology. We have used in our analysis
the following quark masses: $m_u=5$ MeV, $m_d=9$ MeV, $m_s=175$ MeV,
$m_c=1.5$ GeV, $m_b=5$ GeV, $m_t=174$ GeV.

%
\begin{table*}
  \caption{\label{tab:long} Constraints on $\lambda$, $\lambda'$ and
  $\Lambda$ from their contribution to neutrino masses, using recent
  global analysis of the neutrino oscillation data \cite{maltoni}, the
  currently best experimental limit on the $0\nu\beta\beta$-decay
  half-life \cite{H-M}, and matrix element of Ref.\ \cite{FedorVogel}.}
  \begin{ruledtabular}
    \begin{tabular}{cllllll}
      &\multicolumn{2}{c}{the $0\nu\beta\beta$-decay limit} & \multicolumn{2}{c}{normal hierarchy} & \multicolumn{2}{c}{inverted hierarchy}  \\
\cline{2-3} \cline{4-5} \cline{6-7} 
      $m_0=m_{1/2}=$ & 150 GeV & 1000 GeV & 150 GeV  & 1000 GeV & 150 GeV  & 1000 GeV \\
       \hline
      $|\Lambda_e |^2$ [GeV$^2$]    & $1.7 \times 10^{-2}$ & $5.3 $ & $9.0 \times 10^{-5}$  & $2.5 \times 10^{-2}$  & $1.5 \times 10^{-3}$ & $4.6 \times 10^{-1}$ \\
      $|\Lambda_\mu |^2$ [GeV$^2$]  & $1.7 \times 10^{-2}$ & $5.3 $ & $9.1 \times 10^{-4}$  & $2.7 \times 10^{-2}$  & $7.7 \times 10^{-4}$ & $2.3 \times 10^{-1}$ \\
      $|\Lambda_\tau |^2$ [GeV$^2$] & $1.7 \times 10^{-2}$ & $5.3 $ & $8.8 \times 10^{-4}$  & $2.6 \times 10^{-2}$  & $8.4 \times 10^{-4}$ & $2.5 \times 10^{-1}$ \\
      \\
      $\lambda'_{111} \lambda'_{111}$ & $3.2 \times 10^{-3}$ & $2.4 \times 10^{-2}$ & $1.6 \times 10^{-5}$  & $1.2 \times 10^{-4}$  & $2.7 \times 10^{-4}$ & $2.0 \times 10^{-3}$ \\
      $\lambda'_{122} \lambda'_{122}$ & $8.4 \times 10^{-6}$ & $6.2 \times 10^{-5}$ & $4.1 \times 10^{-8}$  & $3.1 \times 10^{-7}$  & $7.2 \times 10^{-7}$ & $5.3 \times 10^{-6}$ \\
      $\lambda_{122}  \lambda_{122}$  & $1.5 \times 10^{-5}$ & $1.1 \times 10^{-4}$ & $7.2 \times 10^{-8}$  & $5.2 \times 10^{-7}$  & $1.2 \times 10^{-6}$ & $9.0 \times 10^{-6}$ \\
      $\lambda'_{133} \lambda'_{133}$ & $8.5 \times 10^{-9}$ & $6.0 \times 10^{-8}$ & $4.2 \times 10^{-11}$ & $3.0 \times 10^{-10}$ & $7.3 \times 10^{-10}$& $5.1 \times 10^{-9}$ \\
      $\lambda_{133}  \lambda_{133}$  & $4.3 \times 10^{-8}$ & $3.1 \times 10^{-7}$ & $2.1 \times 10^{-10}$ & $1.5 \times 10^{-9}$  & $3.7 \times 10^{-9}$ & $2.6 \times 10^{-8}$ \\
      $\lambda'_{132} \lambda'_{123}$ & $2.6 \times 10^{-7}$ & $1.9 \times 10^{-6}$ & $1.3 \times 10^{-9}$  & $9.5 \times 10^{-9}$  & $2.2 \times 10^{-8}$ & $1.6 \times 10^{-7}$ \\
      $\lambda_{132}  \lambda_{123}$  & $8.0 \times 10^{-7}$ & $5.8 \times 10^{-6}$ & $3.9 \times 10^{-9}$  & $2.8 \times 10^{-8}$  & $6.8 \times 10^{-8}$ & $4.9 \times 10^{-7}$ \\
      \\			                                                  												
      $\lambda'_{133} \lambda'_{233}$ & $1.1 \times 10^{-8}$ & $7.7 \times 10^{-8}$ & $7.6 \times 10^{-11}$ & $5.3 \times 10^{-10}$ & $4.8 \times 10^{-10}$& $3.4 \times 10^{-9}$ \\
      $\lambda'_{132} \lambda'_{223}$ & $3.3 \times 10^{-7}$ & $2.5 \times 10^{-6}$ & $2.3 \times 10^{-9}$  & $1.7 \times 10^{-8}$  & $1.5 \times 10^{-8}$ & $1.0 \times 10^{-7}$ \\
      $\lambda'_{123} \lambda'_{232}$ & $3.3 \times 10^{-7}$ & $2.5 \times 10^{-6}$ & $2.3 \times 10^{-9}$  & $1.7 \times 10^{-8}$  & $1.5 \times 10^{-8}$ & $1.0 \times 10^{-7}$ \\
      $\lambda'_{122} \lambda'_{222}$ & $1.1 \times 10^{-5}$ & $8.1 \times 10^{-5}$ & $7.5 \times 10^{-8}$  & $5.6 \times 10^{-7}$  & $4.7 \times 10^{-7}$ & $3.5 \times 10^{-6}$ \\
      $\lambda_{133}  \lambda_{233}$  & $5.5 \times 10^{-8}$ & $4.0 \times 10^{-7}$ & $8.5 \times 10^{-10}$ & $2.8 \times 10^{-9}$  & $2.4 \times 10^{-9}$ & $1.7 \times 10^{-8}$ \\
      $\lambda_{123}  \lambda_{232}$  & $1.0 \times 10^{-6}$ & $7.4 \times 10^{-6}$ & $7.1 \times 10^{-9}$  & $5.1 \times 10^{-8}$  & $4.5 \times 10^{-8}$ & $3.2 \times 10^{-7}$ \\
      \\			                                                  												
      $\lambda'_{233} \lambda'_{233}$ & $1.0 \times 10^{-8}$ & $7.1 \times 10^{-8}$ & $4.2 \times 10^{-10}$ & $3.0 \times 10^{-9}$  & $3.5 \times 10^{-10}$& $2.5 \times 10^{-9}$ \\
      $\lambda'_{232} \lambda'_{223}$ & $3.1 \times 10^{-7}$ & $2.3 \times 10^{-6}$ & $1.2 \times 10^{-8}$  & $9.5 \times 10^{-8}$  & $1.1 \times 10^{-8}$ & $8.0 \times 10^{-8}$ \\
      $\lambda'_{222} \lambda'_{222}$ & $9.9 \times 10^{-6}$ & $7.4 \times 10^{-5}$ & $4.2 \times 10^{-7}$  & $3.1 \times 10^{-6}$  & $3.5 \times 10^{-7}$ & $2.6 \times 10^{-6}$ \\
      $\lambda_{233}  \lambda_{233}$  & $5.1 \times 10^{-8}$ & $3.7 \times 10^{-7}$ & $2.1 \times 10^{-9}$  & $1.5 \times 10^{-8}$  & $1.8 \times 10^{-9}$ & $1.3 \times 10^{-8}$ \\
      \\			                                                  												
      $\lambda'_{133} \lambda'_{333}$ & $1.1 \times 10^{-8}$ & $7.6 \times 10^{-8}$ & $8.0 \times 10^{-11}$ & $5.6 \times 10^{-10}$ & $4.7 \times 10^{-10}$& $3.3 \times 10^{-9}$ \\
      $\lambda'_{132} \lambda'_{323}$ & $3.3 \times 10^{-7}$ & $2.4 \times 10^{-6}$ & $2.4 \times 10^{-9}$  & $1.8 \times 10^{-8}$  & $1.4 \times 10^{-8}$ & $1.0 \times 10^{-7}$ \\
      $\lambda'_{123} \lambda'_{332}$ & $3.3 \times 10^{-7}$ & $2.4 \times 10^{-6}$ & $2.4 \times 10^{-9}$  & $1.8 \times 10^{-8}$  & $1.4 \times 10^{-8}$ & $1.0 \times 10^{-7}$ \\
      $\lambda'_{122} \lambda'_{322}$ & $1.1 \times 10^{-5}$ & $8.0 \times 10^{-5}$ & $7.9 \times 10^{-8}$  & $5.9 \times 10^{-7}$  & $4.6 \times 10^{-7}$ & $3.4 \times 10^{-6}$ \\
      $\lambda_{132}  \lambda_{323}$  & $1.0 \times 10^{-6}$ & $7.3 \times 10^{-6}$ & $7.5 \times 10^{-9}$  & $5.4 \times 10^{-8}$  & $4.4 \times 10^{-8}$ & $3.2 \times 10^{-7}$ \\
      $\lambda_{123}  \lambda_{322}$  & $1.8 \times 10^{-5}$ & $1.3 \times 10^{-4}$ & $1.4 \times 10^{-7}$  & $1.0 \times 10^{-6}$  & $8.1 \times 10^{-7}$ & $5.9 \times 10^{-6}$ \\
      \\			                                                  												
      $\lambda'_{233} \lambda'_{333}$ & $1.1 \times 10^{-8}$ & $7.6 \times 10^{-8}$ & $4.1 \times 10^{-10}$ & $2.9 \times 10^{-9}$  & $3.6 \times 10^{-10}$& $2.6 \times 10^{-9}$ \\
      $\lambda'_{232} \lambda'_{323}$ & $3.3 \times 10^{-7}$ & $2.4 \times 10^{-6}$ & $1.2 \times 10^{-8}$  & $9.3 \times 10^{-8}$  & $1.1 \times 10^{-8}$ & $8.3 \times 10^{-8}$ \\
      $\lambda'_{223} \lambda'_{332}$ & $3.3 \times 10^{-7}$ & $2.4 \times 10^{-6}$ & $1.2 \times 10^{-8}$  & $9.3 \times 10^{-8}$  & $1.1 \times 10^{-8}$ & $8.3 \times 10^{-8}$ \\
      $\lambda'_{222} \lambda'_{322}$ & $1.1 \times 10^{-5}$ & $8.0 \times 10^{-5}$ & $4.1 \times 10^{-7}$  & $3.0 \times 10^{-6}$  & $3.6 \times 10^{-7}$ & $2.7 \times 10^{-6}$ \\
      $\lambda_{232}  \lambda_{323}$  & $1.0 \times 10^{-6}$ & $7.3 \times 10^{-6}$ & $3.8 \times 10^{-8}$  & $2.7 \times 10^{-7}$  & $3.4 \times 10^{-8}$ & $2.5 \times 10^{-7}$ \\
      \\			                                                  												
      $\lambda'_{333} \lambda'_{333}$ & $1.2 \times 10^{-8}$ & $8.3 \times 10^{-8}$ & $4.0 \times 10^{-10}$ & $2.8 \times 10^{-9}$  & $3.9 \times 10^{-10}$& $2.7 \times 10^{-9}$ \\
      $\lambda'_{332} \lambda'_{323}$ & $3.6 \times 10^{-7}$ & $2.6 \times 10^{-6}$ & $1.2 \times 10^{-8}$  & $9.1 \times 10^{-8}$  & $1.2 \times 10^{-8}$ & $8.7 \times 10^{-8}$ \\
      $\lambda'_{322} \lambda'_{322}$ & $1.1 \times 10^{-5}$ & $8.6 \times 10^{-5}$ & $4.0 \times 10^{-7}$  & $3.0 \times 10^{-6}$  & $3.8 \times 10^{-7}$ & $2.8 \times 10^{-6}$ \\
      $\lambda_{322}  \lambda_{322}$  & $2.0 \times 10^{-5}$ & $1.4 \times 10^{-4}$ & $7.0 \times 10^{-7}$  & $5.0 \times 10^{-6}$  & $6.7 \times 10^{-7}$ & $4.8 \times 10^{-6}$ \\
    \end{tabular}
  \end{ruledtabular}
\end{table*}
%

Table \ref{tab:long} shows improved upper bounds on various combinations
of coupling constants of the $\lambda$, $\lambda'$ and $\Lambda$ types,
to be compared with the limits presented in
Refs. \cite{Bhatta,abada}. The results were obtained for $A_0~
=~500~\GeV$, $m_0~=~m_{1/2}~=~150~\GeV$ and $1000~\GeV$ and for positive
$\mu$. We have found a weak dependence of the quantities under
discussion on $A_0$ SUSY parameter. Besides, we have kept $\tan \beta$
large ($\tan \beta \approx 20$), leaving the discussion of the impact of
this parameter on the results to the forthcoming paper. In general, the
new bounds related to lower limit on the $T^{0\nu}_{1/2}$($^{76}$Ge) and
neutrino oscillation data are at least one order of magnitude stronger
than those previously given \cite{Bhatta}. It is mostly due to the
assumption of the gravity-mediated (SUGRA) supersymmetry breaking and
partially due to an improved treatment of the squark and slepton
mixing. As expected the values of R-parity violating coupling related to
normal hierarchy of neutrino masses are significantly suppressed in
comparison with those related to the current lower limit on the
$0\nu\beta\beta$-decay half-life \cite{H-M}.

The new bounds are surprisingly close to those published in
Ref. \cite{abada}, although the method used by the authors of these
papers relayed on many simplifying assumptions. In particular, it
involves setting some of the couplings to zero and assuming all other to
be of the same order of magnitude. Also the whole mechanism of RGE
running as well as GUT constraints were not used. In general the
constraints in \cite{abada} were $\lambda_{x33},\lambda'_{x33}\le
10^{-8}$ which is fully consistent with our results. The bounds on
products of individual coupling constants in Tab. \ref{tab:long} are
either of the same order of magnitude or 1--3 orders of magnitude
stronger.

A more optimistic scenario appears for the case of inverted
hierarchy. In general the corresponding values of product of $\lambda$
and $\lambda'$ coupling are about by factor four less stringent as those
associated with the most stringent $0\nu\beta\beta$-decay limit on the
half-life. We stress again that these values of the R-parity breaking
parameters were determined by the condition that a particular R-parity
breaking mechanism dominates at a time. In some cases this might be
excluded by the phenomenology of other processes. For example, from the
R-parity breaking SUSY mechanism of the $0\nu\beta\beta$-decay one gets
the upper limit on the parameter $\lambda'_{111}$ of the order of
$10^{-4}$ \cite{wodprc99,onbb111}, what is significantly less than the
value presented in Table \ref{tab:long}.

In summary, we have used the GUT constrained R-parity violating minimal
supersymmetric standard model to describe massive neutrinos. The three
family neutrino mass matrix was calculated within framework including
the tree-level neutrino-neutralino mixing and the one-loop radiative
corrections. Then, the theoretical mass matrix was compared with the
phenomenological one, obtained by using the most recent global analysis
of neutrino oscillations data and the lower limit on the half-life of
neutrinoless double beta decay of $^{76}$Ge. This procedure allowed
to improve the upper limits on certain products of R-parity violating
couplings, which are up to one order of magnitude more stringent as
those previously published \cite{Bhatta}. Further on, we assumed the
normal and inverted hierarchy of neutrino masses and calculated the
corresponding values of R-parity violating parameters of the SUSY model
under consideration. These values can be used in determining the
perspectives of finding signal of R-parity violation in different
experiments, in particular at colliders. This issue is, however, beyond
the scope of this paper.

{\bf Acknowledgments} This work was supported by the VEGA Grant agency
of the Slovak Republic under contract No.~1/0249/03, by the EU ILIAS
project under contract RII3-CT-2004-506222, and the Polish State
Committee for Scientific Research under grants no. 2P03B~071~25 and
1P03B~098~27.



\end{document}